# Enhancing Mars Life Explorer (MLE) with True Agnostic Life Detection Capabilities

Gabriella Rizzo[1,2], Jan Spacek[1,3] (info@alfamars.org)

[1]Agnostic Life Finding Association Inc., Alachua, FL, USA; [2]School of Biological Sciences, University of Nebraska-Lincoln, Lincoln, NE, USA; [3]Foundation for Applied Molecular Evolution, Alachua, FL, USA

## 1. Introduction

The Mars Life Explorer (MLE) mission concept represents a key opportunity to assess whether extant life exists within the mid-latitude ice deposits of Mars. However, MLE's current instrument suite, as defined in the Mars Life Explorer Science Traceability Matrix, is optimized for evaluating habitability and organic chemistry, *not for directly detecting extant life* [1]. With human missions to Mars potentially occurring as early as 2040, the time window for uncontaminated robotic investigation is narrowing. A high-confidence assessment of the Martian biosphere must occur before anthropogenic contamination negatively affects life detection. This paper identifies key scientific, technical, and policy limitations in the current MLE framework and proposes specific upgrades and governance strategies to ensure robust life detection and planetary protection compliance.

## 2. Constraints in Current Life Detection Strategy

### 2.1 Legacy of Viking

The 1976 Viking landers remain the only missions to explicitly test for extant life on Mars. While their biological experiments returned ambiguous or null results, multiple reanalyses have shown that the failure to detect organics was due to method limitation, *not because of their absence in the sample*. Non-volatile organic compounds such as mellitic acid or oxalate, potentially derived from meteoritic infall, could have escaped detection by Viking's GC-MS due to thermal instability or poor volatility [2]. Other studies suggest that Viking's results can be explained by perchlorate chemistry and radiation-driven oxygen release [3], without invoking reactive superoxides or extant life, though the biological interpretation of the Labeled Release experiment remains unresolved and continues to be defended by some as evidence for microbial metabolism [4]. Regardless, these results highlight that *the absence of detectable biosignatures under constrained methodologies does not imply the absence of life*.

### 2.2 Narrow Molecular Target Space in MLE

MLE currently targets a narrow suite of molecules such as amino acids, fatty acids, polynucleotides, and PAHs, that are consistent with terrestrial biology but not diagnostic of life in isolation [1]. Each can form abiotically under plausible prebiotic or geochemical conditions [5]. Moreover, the detection thresholds and contextual analyses proposed e.g., bulk concentration measurements without enantiomeric excess, isotopic fractionation, or spatial organization, are unlikely to unambiguously distinguish biological origin from abiotic processes.

### 2.3 Absence of Sensitive Agnostic Detection Capabilities

Importantly, MLE does not include any agnostic life detection technologies capable of evaluating features beyond Earth-centric biochemistry such as detection of (alien) genetic polymers, informational complexity, or molecular organization. This limitation assumes, implicitly, that Martian life (if present) would fall within the detection space of known terrestrial biomolecules. However, research in ultra-low-biomass ecosystems on Earth, such as hyper-arid deserts and high-UV environments, has shown that instruments traditionally used in astrobiology often lack the sensitivity needed to detect life under such extreme conditions [6]. If Martian biology exists in similarly low concentrations, the current MLE framework may be not sensitive enough to find it, even if Martian life does not differ in its biochemistry from Earth's. MLE, as proposed, is likely to fail to find life with unknown biochemistry.

## 3. Temporal and Planetary Protection Constraints

### 3.1 Mission Timing and Scientific Ambiguity

NASA, CNSA, and private industry have expressed interest in launching crewed Mars missions by the 2040s. A full MLE mission lifecycle, including instrument finalization, integration, launch, cruise, surface operations, and data analysis, could extend into the late 2030s. Any significant delay, technical issue, or ambiguity in MLE's findings would not leave sufficient time for a follow-up robotic investigation before human arrival. As MLE Science Champion Amy Williams acknowledged, "NASA or commercial partners might have to send another mission to further investigate", since none of the proposed instruments can provide a conclusive detection

[7]. This reinforces the concern that the last uncontaminated window for detecting extant life could close without yielding a definitive result.

## 3.2 Risk of Irreversible Contamination

This timeline poses a planetary protection risk. Human-associated microbes have been recovered from spacecraft and assembly facilities despite rigorous bioburden reduction protocols. Some of these organisms, including spore-forming strains, have demonstrated exceptional resistance to simulated Martian UV conditions, surviving exposure to UVA, UVB, and UVC spectra for hours at intensities comparable to those on the Martian surface [8]. This suggests that, if such Earth-origin organisms were transported to Mars, they could potentially survive long enough to be carried hundreds of miles to more hospitable locations. Though, it is worth noting that UV radiation is not a concern during nighttime or global dust storms. Moreover, many terrestrial microbes, particularly extremophiles, can survive the desiccation, low temperatures, and high salinity characteristic of the Martian near-surface, especially when shielded from UV by regolith or dust cover [9]. These findings challenge assumptions that Mars is uniformly biocidal and underscore the risk that introduced terrestrial organisms could remain viable on Mars for extended periods. If inadvertently introduced into subsurface ice or hydrated environments, these microbes could colonize and confound detection efforts, particularly if Martian life shares even partial biochemical similarities.

## 4. Recommendations for Mission Enhancement
## 4.1 Integration of an Agnostic Life Detecting Instrument, or the Agnostic Life Finder (ALF)

To meet its science objectives and comply with planetary protection standards, MLE should incorporate at least one agnostic life detection system capable of identifying informational polymers without presupposing Earth-like biochemistry. A strong candidate is the Agnostic Life Finder (ALF), designed to concentrate and characterize polyelectrolytes such as DNA, RNA, or non-terrestrial analogs from aqueous environments [10].

ALF uses continuous electrodialysis and size-exclusion membranes to separate and concentrate high-molecular-weight, charged polymers from large water volumes. Its sensitivity scales with sample volume, enabling detection of ultra-trace biopolymers. Known nucleic acids are sequenced via biological nanopores; unknown polymers are analyzed using solid-state nanopores and/or fragmentation mass spectrometry to assess structural regularity (e.g., Schrödinger-type heteropolymer ordering).

ALF advanced to TRL 4 under NASA's NIAC Phase I program, and could further be matured to flight readiness within the MLE development timeline, assuming early prioritization [11]. Its design is compatible with In-Situ Resource Utilization (ISRU)-scale water access systems envisioned for future crewed missions, linking exploration and biosignature detection. MLE's modular payload architecture, along with Redwater Rodwell ability, can accommodate such instrumentation if prioritized early [12].

We also strongly recommend metagenomic methods (with or without preconcentration) to monitor and analyze the forward contamination as well DNA/RNA based Martian life forms.

## 4.2 Establishing a 'Scientific Clearance' Protocol

We also recommend that COSPAR and national space agencies adopt a formal scientific clearance protocol requiring that any crewed Mars landing be preceded by in situ biosignature assessment [13]. This would establish an evidentiary threshold comparable to biosafety containment in terrestrial microbiology, ensuring that life detection is resolved before irreversible anthropogenic contamination occurs. Without such a safeguard, any future claims, positive or negative, about Martian life risk being invalidated by uncertainty over Earth-sourced interference. Analogous clearance thresholds already exist for planetary sample return, high-containment lab work, and other domains where biological ambiguity carries irreversible consequences. Extending this standard to Mars is both scientifically prudent and ethically necessary.

## 5. Conclusion

MLE might be our best opportunity to determine whether life exists beyond Earth, but its current design lacks the tools to deliver a definitive answer. Without the ability to perform a sensitive detection of agnostic biosignatures, MLE risks confusing biological signals with abiotic chemistry or *vice versa*. Once humans arrive on Mars, that ambiguity will be harder to resolve. Upgrading the payload is feasible and necessary. MLE must evolve from a habitability study into a true life detection mission, and the time to act is now.